\documentclass[aps,prb,twocolumn,superscriptaddress,citeautoscript,amssymb,reprint,showpacs,floatfix,longbibliography]{revtex4-2}
\usepackage[utf8]{inputenc}
\usepackage[T1]{fontenc}
\usepackage{float}
\usepackage{amsmath}
\usepackage{ulem}
\usepackage{mathptmx}
\usepackage{mathrsfs}
\usepackage[english]{babel}
\usepackage{graphicx}
\usepackage{color}
\usepackage{overpic}

\usepackage{siunitx}
\newcommand{\SRO}{Sr\textsubscript{2}RuO\textsubscript{4}}

\definecolor{fj_color}{cmyk}{1, 0.3, 0, 0}
\definecolor{jl_color}{cmyk}{0, 0.8, 0.8, 0}
\definecolor{mp_colour}{cmyk}{0.8, 0, 0.8, 0}
\definecolor{ar_colour}{rgb}{0, 1, 0}
\definecolor{movethis}{rgb}{1,0,0}

\begin{document}

\title{Quantum Oscillations of \SRO{} under $c$-Axis Uniaxial Stress}

\author{Fabian Jerzembeck}
%\email{Fabian.Jerzembeck@cpfs.mpg.de}
\affiliation{Max Planck Institute for Chemical Physics of Solids, N\"{o}thnitzer Str 40, 01187 Dresden, Germany}
\author{Maximilian T. Pelly}
\affiliation{Scottish Universities Physics Alliance (SUPA), School of Physics and Astronomy,
University of St Andrews, St Andrews KY16 9SS, United Kingdom}
\author{Helge Rosner}
\affiliation{Max Planck Institute for Chemical Physics of Solids, N\"{o}thnitzer Str 40, 01187 Dresden, Germany}
\author{Edgar Abarca Morales}
\affiliation{Max Planck Institute for Chemical Physics of Solids, N\"{o}thnitzer Str 40, 01187 Dresden, Germany}
\author{Naoki Kikugawa}
\affiliation{National Institute for Materials Science, Tsukuba 305-0003, Japan}
\author{Dmitry A. Sokolov}
\affiliation{Max Planck Institute for Chemical Physics of Solids, N\"{o}thnitzer Str 40, 01187 Dresden, Germany}

\author{Andrew P. Mackenzie}
\affiliation{Max Planck Institute for Chemical Physics of Solids, N\"{o}thnitzer Str 40, 01187 Dresden, Germany}
\affiliation{Scottish Universities Physics Alliance (SUPA), School of Physics and Astronomy,
University of St Andrews, St Andrews KY16 9SS, United Kingdom}

\author{Andreas W. Rost}
\email{a.rost@st-andrews.ac.uk}
\affiliation{Scottish Universities Physics Alliance (SUPA), School of Physics and Astronomy,
University of St Andrews, St Andrews KY16 9SS, United Kingdom}

\author{Elena Hassinger}
\affiliation{Institute for Quantum Materials and Technologies, Karlsruhe Institute of Technology, Kaiserstra\ss e 12, 76131 Karlsruhe, Germany}
\affiliation{Max Planck Institute for Chemical Physics of Solids, N\"{o}thnitzer Str 40, 01187 Dresden, Germany}

\author{Javier F. Landaeta}
\email{Javier.Landaeta@cpfs.mpg.de}
\affiliation{Max Planck Institute for Chemical Physics of Solids, N\"{o}thnitzer Str 40, 01187 Dresden, Germany}

\date{\today}

\begin{abstract}

Uniaxial stress has now been widely used to study correlated electron materials.
However, Fermi surface-resolved experimental data on the evolution of the electronic structure under piezoelectrically applied stress are sparse, with no reports of de Haas-van Alphen (dHvA) effects under uniaxial stress.
Here we present dHvA measurements under $c$-axis uniaxial stress on the unconventional superconductor \SRO{}. This allows us to study the evolution of the electronic structure directly and to gain insight into the contradicting behavior of the predicted enhancement of the electronic density of states and the observed suppression of $T_\text{c}$.
We are able to follow all Fermi surfaces for stress up to $-1.8$~GPa and find that the cross-sectional areas of the hole-like $\alpha$ sheet increase and electron-like $\beta$ sheet decrease.
At the same time, the area of the electron-like $\gamma$ sheet increases.
Therefore, in contrast to in-plane uniaxial stress, charge transfer is the mechanism for approaching the electron-to-hole Lifshitz transition and the associated Van Hove singularity. 
Additionally, we find that the effective masses on all three Fermi sheets are slightly enhanced as the Lifshitz transition is approached.
We compare the dHvA results with quantum oscillations in the magnetostriction and band structure calculations, and find good agreement.
At a more general level, our findings show that quantum oscillation measurements under uniaxial stress, combined with band-structure calculations, offer a promising new route for studying quantum materials.

\end{abstract}
\maketitle

\section{Introduction}

In recent years, uniaxial stress has been widely used to tune quantum materials \cite{Hicks25_ARCMP, Jo24_FEM} and study, for example, the nematic phase in the iron-pnictides \cite{Chu12_Science, Ikeda21_PNAS}, charge density waves in the cuprates \cite{Kim18_Science, Nakata22_npjQM}, or quantum critical systems \cite{Malinowski20_NatPhys, Worasaran21_Science} by electrical transport or magnetic susceptibility measurements. 
However, not much experimental work has been dedicated to quantifying how the effect of uniaxial stress on measured quantities is related to changes in the underlying electronic structure and Fermi surface.

\begin{figure}[ptb]
\includegraphics[width=\columnwidth]{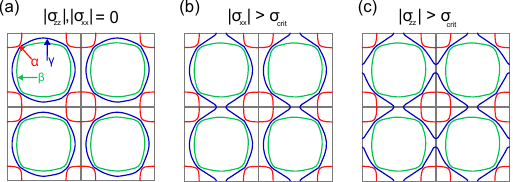}
\caption{Schematic 2D Fermi surface of \SRO{} under uniaxial stress. 
(\textbf{a}) Zero stress. 
(\textbf{b}) In-plane uniaxial stress $|\sigma_{xx}| > |\sigma_{\text{crit},xx}|$. 
(\textbf{c}) Out-of-plane uniaxial stress $|\sigma_{zz}| > |\sigma_{\text{crit},zz}|$.}
\label{fig:FS_schematic}
\end{figure}

ARPES under uniaxial stress was used to study the quasiparticle structure of the iron-pnictides \cite{Pfau19_PRL, Pfau19_PRB} and the unconventional superconductor \SRO{} under $\langle 100 \rangle$ uniaxial stress \cite{Sunko19_npj}.
In the latter, a Lifshitz transition was observed when one of the Fermi sheets (the $\gamma$ sheet) is tuned from a closed (electron-like) to an open Fermi surface as it touches the Brillouin-zone (BZ) boundary (see Fig.~\ref{fig:FS_schematic}(b)). 
By tuning through this Lifshitz transition and an associated Van Hove singularity, the density of states is strongly enhanced, and $T_\text{c}$ and $H_\text{c2}$ peak at a critical stress of $\sigma_\text{crit,a}^\text{exp} \approx -0.7$~GPa (minus denotes compression) \cite{Steppke17_Science, Barber18_PRL, Li22_Science, Jerzembeck23_PRB}. On the other hand, $c$-axis uniaxial stress is predicted to tune to a Lifshitz transition with a Van Hove singularity when the $\gamma$ sheet touches the BZ boundary at multiple points at $\sigma_\text{crit,c} \approx -5.5$~GPa \cite{Jerzembeck22_NatComm}, changing from an electron-like to a hole-like Fermi surface (Figure~\ref{fig:FS_schematic}(c)). 
This Van Hove singularity is expected to enhance the density of states even more. Surprisingly, however, a previous experiment found that $c$-axis uniaxial stress up to $-3.2$~GPa weakly suppresses $T_\text{c}$ while it enhances the upper critical field \cite{Jerzembeck22_NatComm}. 
This behavior is unexpected, since an enhanced density of states would naively be expected to increase $T_\text{c}$ as well, and may therefore provide insight into the still unresolved nature of superconductivity in \SRO{}.

In order to study whether the suppression of the superconducting $T_c$ with $c$-axis uniaxial stress is actually at odds with the predicted evolution of the band structure, we measure the de Haas-van Alphen (dHvA) effect under $c$-axis uniaxial stress. In principle, the de Haas-van Alphen effect is the ideal technique for clean systems, because it allows us to study the changes of this correlated material ($m^*/m_\text{DFT}|_{\alpha, \beta, \gamma} = 3 - 5.5$ \cite{Mackenzie03_RMP}) at low temperatures ($\sim 100$~mK). 
However, when combined with uniaxial stress, quantum oscillation measurements face several obstacles, especially in correlated materials with large mass enhancements. The required low temperatures are challenging to achieve with a piezoelectric-based uniaxial stress cell made of titanium.
Furthermore, strain inhomogeneity was found to affect uniaxial stress studies \cite{Steppke17_Science, Jerzembeck23_PRB} and is also believed to severely impact quantum oscillation measurements. Hence, studies of quantum oscillations under uniaxial stress have so far focused on measurements of semimetals with small effective masses and under comparably small strains, $|\varepsilon| < 0.4~\%$ in transport (Shubnikov-de Haas)~\cite{Jo19_PNAS, Schindler20_PRB}. 

Here, we report dHvA measurements for uniaxial stress up to $-1.8$~GPa ($\varepsilon_{zz} \approx -0.85~\%$) along the $c$-axis of \SRO{}. We find, in agreement with band-structure calculations, that all dHvA frequencies shift with applied $c$-axis uniaxial stress. The areas of the hole-like $\alpha$ and the electron-like $\gamma$-sheet increase, whereas that of the electron-like $\beta$-sheet area decreases, supporting the idea of a charge transfer between the Fermi sheets. 

To complement the dHvA experiments, we also study quantum oscillations in the magnetostriction, using bespoke apparatus designed for the purpose. These magnetostriction oscillations provide complementary results on the stress dependence of the Fermi surface frequencies at zero stress, that is, $\frac{\partial F}{\partial \sigma}\big |_{\sigma = 0}$. Both techniques agree quantitatively with each other and, additionally, qualitatively with three-dimensional density functional theory (DFT) calculations. Furthermore, we determine the effective masses from quantum oscillations and the DFT band masses at a compression of -1 GPa and find that the $c$-axis stress weakly enhances both, leaving the ratio constant within uncertainties.

%%%%%%%%%%%%%%%%%%%%%%%%%%%%%%%%%%%%%%%%
\section{Experimental results}

\subsection{dHvA Measurements}
\label{sec:dHvA}

\begin{figure*}[ptb]
\includegraphics{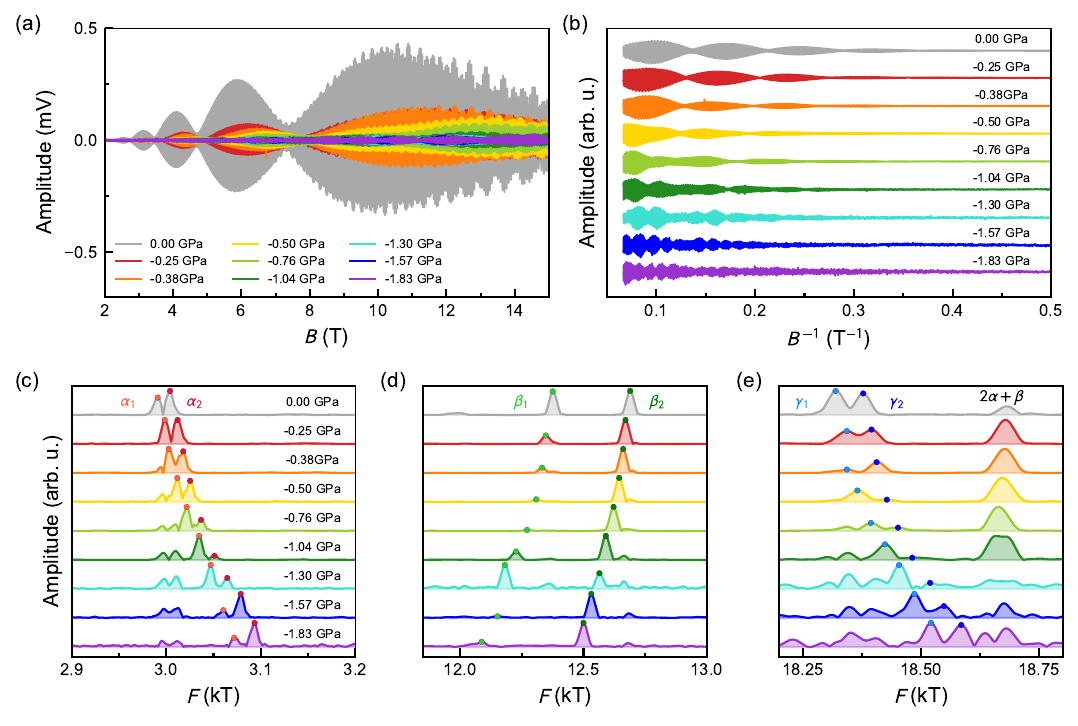}
\caption{(\textbf{a}) Raw dHvA oscillations as a function of magnetic field at 90~mK (stress cell temperature) for a series of compressive stresses. 
(\textbf{b}) Same data as in (\textbf{a}) plotted versus inverse field; curves are vertically shifted for clarity. 
(\textbf{c--e}) Fourier spectra highlighting the $\alpha$, $\beta$, $\gamma$, and $2\alpha+\beta$ peaks under different stresses. FFTs were taken over the field range 3--15~T for $\alpha$ and 8--15~T for the other Fermi surface sheets.}
\label{fig:raw-QO}
\end{figure*}

To study the dHvA quantum oscillations under uniaxial stress, we used a field-modulation technique and a uniaxial stress cell based on piezoelectric actuators \cite{Hicks14_RSI, Barber19_RSI}, equipped with a displacement (strain) sensor. 
The field-modulation technique used a pick-up coil with 48 turns and an excitation coil wound on a hollow cylinder with 100 turns. 
In our measurements, we first determined the stress scale by measuring $T_\text{c}$ as a function of applied displacement and converting displacement to stress using a method introduced in Ref.~\cite{Jerzembeck22_NatComm}. More details on the setup and the stress conversion can be found in the Appendix. 
Since the sample is free to undergo in-plane deformation, applying $c$-axis stress induces, by symmetry, biaxial in-plane tensile strain in addition to the compressive strain along the $c$ axis. For the subsequent measurements of the dHvA effect, we maintained a stress level lower than the maximum applied stress during the $T_c$ study. 
Figure~\ref{fig:raw-QO}(a) shows raw quantum oscillations at base temperature (90 mK on the stress cell thermometer) for a series of compressive stresses as a function of applied magnetic field.

Firstly, we observe that as stress increases, the amplitude of the quantum oscillations decreases, which we attribute to strain inhomogeneity. Nevertheless, clear oscillations remain visible up to the largest applied stress of $-1.83$~GPa.
Secondly, the beating pattern of the oscillations changes with applied stress, which can be better seen in Fig.~\ref{fig:raw-QO}(b), where the normalized oscillations are plotted against inverse field and shifted for clarity. The clear change in the beating pattern already demonstrates that the electronic structure of \SRO{} is tuned by stress. The warping of the Fermi surface, which produces the observed beating in the oscillation amplitude, becomes more pronounced with increasing stress as can be deduced from the observed beat patterns in Fig.~\ref{fig:raw-QO}. These clear quantum oscillations allow us to determine the dHvA frequencies of the $\alpha$-, $\beta$-, and $\gamma$-sheets for stresses of up to $-1.8$~GPa, as shown in Fig.~\ref{fig:raw-QO}(c-e). The zero-stress frequencies are in good agreement with previous quantum oscillation studies \cite{Mackenzie96_PRL, Bergemann03_AIP, Forsythe02_PRL}. Both the $\alpha$ and $\beta$ peaks shift with increasing stress to larger and smaller frequencies, respectively. At large stresses, we observe small peaks close to the zero-stress frequencies (most visible for the $\alpha$ sheet), indicating that we also detect the almost unstrained ends of the sample. This is consistent with our measurements of the temperature dependence of mutual inductance, which we use to detect the diamagnetic signal at the superconducting transition. As shown in the Appendix, a double transition appears at large stresses: a small transition from the almost unstrained ends of the sample and a larger transition from the strained central region. 
%Similarly, at small stresses, the $\gamma$ sheet exhibits a clear shift to larger frequencies. For large stresses, the signal-to-noise ratio decreases notably as a result of the suppression of the dHvA amplitude combined with the large band masses of the $\gamma$ sheet. 
%Despite a decrease in the signal-to-noise ratio at large stresses, our setup allows us to track the evolution of the Fermi sheet up to the largest applied compression. 
Finally, the peak around $18.7$~kT, associated with the $2\alpha + \beta$ orbit, shifts within our resolution like the sum of $2\alpha_1$ and $\beta_2$, further confirming our Fermi-surface identification.

\begin{figure}[ptb]
\includegraphics{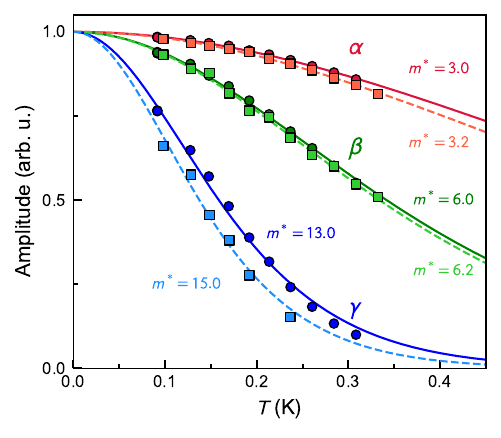}
\caption{Lifshitz-Kosevich effective mass analysis at 0~GPa 
(circles, solid lines) and $-1$~GPa (squares, dashed lines). 
Uncertainties on the effective masses from the fits are estimated to be about 2\% for the $\alpha$ and $\beta$ sheets, 
and about 5\% for the $\gamma$ sheet.}
\label{fig:mass_analysis}
\end{figure}

Next, we determine the effective masses at zero stress and at a compression of $-1$~GPa from the temperature dependence of the quantum-oscillation amplitudes shown in Fig.~\ref{fig:mass_analysis}. At zero stress, the Lifshitz-Kosevich (LK) fits yield effective masses for the three Fermi surfaces that are slightly smaller (by about 10--20\%) than previously reported \cite{Mackenzie96_PRL, Bergemann03_AIP}, consistent with a small heating of the sample relative to the thermometer (see mass analysis section and Fig.~\ref{fig:App_osc} in the Appendix). 
Since the experiment at $-1$~GPa was performed under identical experimental conditions as the zero-stress measurement (apart from the applied DC piezo voltage), the heating effect is expected to be comparable in both cases. We therefore extract the effective masses directly from LK fits to the measured amplitudes, without applying any temperature correction, assuming that the observed relative changes in the effective masses are intrinsic.

At an applied stress of $-1$~GPa, the effective masses of all three Fermi sheets are enhanced on average by $\sim 10\%$ relative to their zero-stress values.
%, consistently across different analysis approaches. 
Despite the reduced oscillation amplitude at high stress, this enhancement is observed for all three sheets. A similar weak enhancement of the effective masses was observed for epitaxially strained films of \SRO{}~\cite{Burganov16_PRL}.
%%%%%

In summary, we have demonstrated that de Haas-van Alphen measurements on a correlated material under $c$-axis uniaxial stress are feasible up to stresses exceeding $-1.8$~GPa. Strain inhomogeneity does occur, but it can be controlled with careful sample preparation and a small mutual inductance setup. This allowed us to measure both the frequency shift and the mass enhancement of the $\alpha$, $\beta$ and $\gamma$ sheet.

%%%%%%%%%%%%%%%%%%%%%%%%%%%%%%%%%%
\subsection{Quantum Oscillations in Magnetostriction}
A complementary measurement of the stress sensitivity of the Fermi surface sheets can be made at zero applied stress using quantum oscillations in sample strain~\cite{Griessen1972a, Griessen1972b, Griessen1974a, Griessen1974b, Griessen1977}. These strain oscillations are related to oscillations in magnetization oscillations via~\cite{Shoenberg1984a}:
\begin{equation}
    \tilde{\varepsilon}_{n,ik} = -\frac{1}{F_n} \frac{\partial F_n}{\partial \sigma_{ik}} \tilde{M}_n H.
    \label{eq:strain_osc}
\end{equation}
Here $\tilde{\varepsilon}_{n,ik}$ is the oscillation in strain of the $n$-th frequency $F_n$, and $\tilde{M}_n$ the corresponding oscillation in magnetisation. 

Voltage oscillations $\tilde{v}_n$ in a dHvA experiment (with field-modulation technique) are given by~\cite{Shoenberg1984b}:
\begin{align}
\begin{split}
    \tilde{v}_n &= -2c\omega J_1\left(\lambda_n\right) \tilde{M}_n\\
    \\
    \lambda_n &= \frac{2\pi F_n h_0}{H^2} \equiv \frac{2\pi h_0}{\Delta H_n}\;,
    \label{eq:dvha_voltage}
    \end{split}
\end{align}

with $c$ being the coupling constant of the pickup coils, $\omega$ and $h_0$ the frequency and amplitude of the AC excitation respectively,  $J_1$ the Bessel function of the first kind and $\Delta H_n$ the field interval of one dHvA oscillation. 

Experimentally we operate in the regime $\lambda_n\ll1$ in which case  the Bessel function reduces to $J_1(\lambda_n) \approx \lambda_n/2$. In this limit equation (\ref{eq:dvha_voltage}) can be expressed as:

\begin{equation}
    \tilde{v}_n = -c\omega \lambda_n \tilde{M}_n = -c\omega \frac{2\pi F_n h_0}{H^2} \tilde{M}_n \;.
    \label{eq:dvha_voltage2}
\end{equation}

Combining equations (\ref{eq:strain_osc}) and (\ref{eq:dvha_voltage2}) allows one to express the strain dependence of an individual frequency as:

\begin{equation}
    \frac{\partial F_n}{\partial \sigma_{ik}} = \tilde{\varepsilon}_{n,ik}\frac{F_n}{\tilde{M}_n H}=a\frac{F_n^2\tilde{\varepsilon}_{n,ik}}{\tilde{v}_n }
    \label{eq:sensitivity0}
\end{equation}
with $a=c\omega 2\pi h_0 / H^3$ containing all measurement parameters independent of the the quantum oscillation frequency.  

\begin{figure}[ptb]
\includegraphics{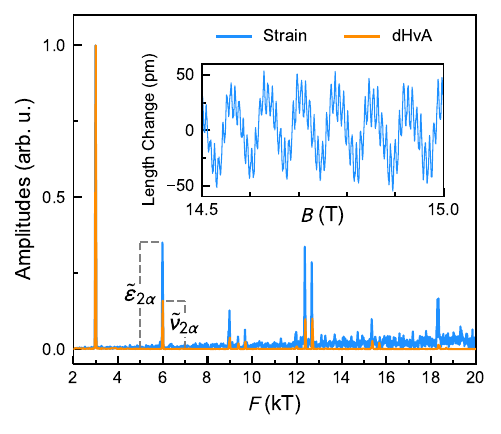}
\caption{Comparison of the FFTs ($8$--$15$~T range) of quantum oscillations 
from magnetostriction (\qty{85}{\milli\kelvin}) and dHvA (\qty{125}{\milli\kelvin}) on the same sample. Before comparison, the strain FFT was multiplied by \(F^2\) (details are in the main text). Both spectra are then normalized to the $\alpha$ peak height. The inset shows the raw dilatometer signal (strain oscillations) as a function of field.}
%old%\caption{Comparison of the FFTs ($8$--$15$~T range) of quantum oscillations 
%from magnetostriction (\qty{85}{\milli\kelvin}) and by dHvA (\qty{125}{\milli\kelvin}) on the same sample. 
%Before comparison, the strain FFT was multiplied by \(F^2\) (details are in the main text). 
%Both spectra are then normalized to the $\alpha$ peak height. The main panel highlights the amplitudes $\tilde{\epsilon}_{2\alpha}$ and $\tilde{v}_{2\alpha}$ which, after correction for the different LK thermal damping factors, are used to compute \(\Lambda_{2\alpha}\). The inset shows the raw dilatometer signal (strain oscillations) as a function of field.}
\label{fig:qo_fft_comparison}
\end{figure}

To measure strain oscillations, we have developed a new capacitive dilatometer based on an Attocube\texttrademark{} linear actuator (see Appendix for details). In order to directly compare strain and dHvA oscillation amplitudes, we studied a 500~$\mu$m-long section of the same sample the dHvA measurements under strain were taken on. In the inset of Figure~\ref{fig:qo_fft_comparison} we show a representative measurement of the strain oscillations between at magnetic fields between 14.5 and \qty{15}{\tesla} at a temperature of \qty{85}{\milli\kelvin}.

Equation (\ref{eq:sensitivity0}) implies a convenient visual approach providing a qualitative indication as to the relative stress dependencies of the quantum oscillation frequencies, which we show in Fig. \ref{fig:qo_fft_comparison}. With $\mathcal{F}$ denoting the FFT, we are plotting $\mathcal{F}\left(\tilde{v}\right)$ as well as $F^2\mathcal{F}\left(\tilde{\varepsilon}\right)$ as a function of frequency $F$.
As the coil constant $c$ is unknown, both datasets are normalized by the respective amplitude of the peak at $F_\alpha$. The stress sensitivities of individual frequencies relative to the stress sensitivity of $F_\alpha$ can now be read off this plot as the ratio of the amplitudes in this plot. 

An important example, that provides a self-consistency check, is given by higher harmonics of a primary frequency, as the stress sensitivity of the \(n^{th}\) harmonic must be \(n\) times that of the fundamental. To be explicit, the frequency of the second harmonic of $F_\alpha$ must change twice as fast with stress compared to first harmonic ($\partial F_{2\alpha}/\partial\sigma = 2\partial F_\alpha/\partial\sigma$). Hence, in the type of plot shown in Fig. \ref{fig:qo_fft_comparison}, the strain related peak at $F_{2\alpha}\approx 6\;\mathrm{kT}$ should be twice as large as the dHvA peak at the same frequency, as is indeed the case.

Equation (\ref{eq:sensitivity0}) is only valid if both datasets are taken at the same temperature. 
For the measurements discussed above, this condition is only approximately fulfilled. The constraint can be overcome by using the zero temperature amplitudes $\tilde{\epsilon}_{n, 0K}$ and $\tilde{v}_{n, 0K}$ based on the Lifshitz-Kosevich temperature dependence, which has been done for the subsequent quantitative analysis.

%Revised up to here , but its need it to discuss with Andreas
%%%%%%%%%%%%%%%%%%%%%%%%%%%%%%%%%%%%
\section{Discussion}
%%%A
%We have shown that the stress dependence of the Fermi surfaces in a correlated material under $c$-axis uniaxial stress can be measured using dHvA and magnetostriction measurements.

\begin{figure}[ptb]
\includegraphics{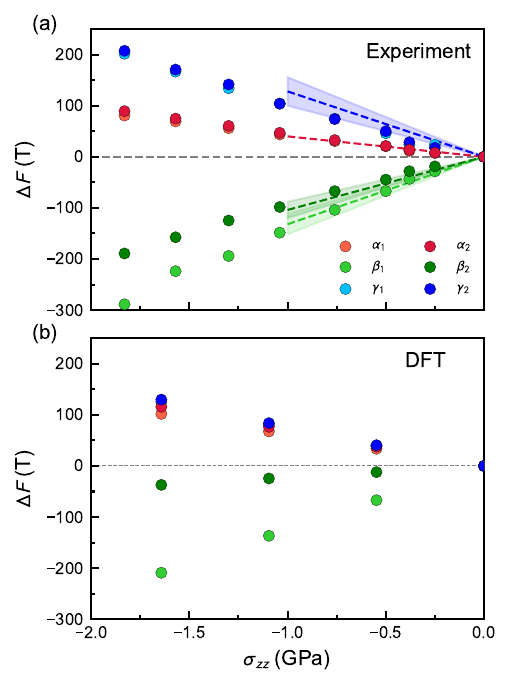}
\caption{(\textbf{a}) Frequency shift $\Delta F = F_n - F_{n,0}$ as a function of stress for all Fermi sheets. Error bars are smaller than the symbol size. The dashed lines and shaded areas indicate slopes and uncertainties obtained from magnetostriction, normalized to the $\alpha$ slope. (\textbf{b}) $\Delta F = F_n - F_{n,0}$ as a function of stress calculated within DFT.}
\label{fig:main_results}
\end{figure}

Here, we directly compare the two measurement techniques that can yield complementary information. We also relate them to DFT calculations. Figure~\ref{fig:main_results}(a) shows the change in the peak frequencies, $\Delta F = F_n - F_{n,0}$, of the $\alpha$, $\beta$, and $\gamma$ Fermi sheets as a function of $c$-axis stress, where $F_{n,0}$ denotes the zero-stress frequencies determined in Fig.~\ref{fig:raw-QO}(c-e). Linear fits for $\sigma_{zz} \geq -0.5$~GPa yield slopes $dF_n/d\sigma_{zz}$ of approximately $-40$~T/GPa and $-84$~T/GPa for the $\alpha_{1,2}$ and $\gamma_{1,2}$ peaks, respectively, and $132$~T/GPa and $86$~T/GPa for the $\beta_1$ and $\beta_2$ peaks. We compare these stress dependencies of the Fermi surface frequencies with the slopes $\partial F_n / \partial \sigma_{zz}\big|_{\sigma_{zz} = 0}$ obtained from magnetostriction measurements. Using Equation (\ref{eq:sensitivity0}), we determine the stress sensitivity of the $n$-th sheet relative to the $\alpha$ sheet by comparing the corresponding zero temperature peak amplitudes. 

The following relation describes this comparison:

\begin{equation}
    \Lambda_n =\frac{\tilde{\varepsilon}_{n,0K}/\tilde{v}_{n,0K}}{\tilde{\varepsilon}_{\alpha,0K}/\tilde{v}_{\alpha,0K}}.
\end{equation}

In Fig.~\ref{fig:main_results}(a), we plot the zero-stress magnetostriction slopes, expressed in units of the $\alpha$ slope ($\Lambda_n \times 40$~T/GPa), as dashed lines. 
The shaded areas around the dashed lines indicate the uncertainties of the magnetostriction measurements. By design, the $\alpha$ sheet agrees exactly, as it is used for normalization. For $\beta_1$ and $\beta_2$, we also find good quantitative agreement between both techniques. In both cases, $\beta_1$ shows a significantly larger stress dependence than $\beta_2$, consistent with the $\beta$ Fermi surface becoming less two-dimensional under $c$-axis stress. Within the uncertainties, the slope of the $\gamma$ sheet also agrees with the value extracted from the dHvA experiment.

Table~\ref{tbl:predictions} summarizes the dHvA slopes and magnetostriction stress sensitivities shown in Fig.~\ref{fig:main_results}(a).

%%%B
As a direct comparison, we calculated the evolution of the Fermi surfaces under $c$-axis strain, $\varepsilon_{zz}$, using density functional theory (DFT). 
The calculations were carried out in the local density approximation (LDA) and used the relevant low-temperature Poisson's ratios and lattice parameters; details are given in the Appendix.
We find 9 extremal orbitals in the first Brillouin zone; 3 for $\alpha$, 2 for $\beta$ and 4 for the $\gamma$ sheet, where two of the $\alpha$ and two of the $\gamma$ orbitals have degenerate frequencies by symmetry.
In agreement with the dHvA measurements, the DFT calculations find that the Fermi surface frequencies of the $\alpha$ and $\gamma$ sheets increase, while those of the $\beta$ sheet decrease.
%%%C
Figure~\ref{fig:main_results}(b) shows the change $\Delta F = F_n - F_{n,0}$ of the extremal orbit frequencies of the DFT Fermi surfaces as a function of $c$-axis stress, using $\sigma_{zz} = E_{zz}\,\varepsilon_{zz}^\text{DFT}$ with $E_{zz} = 219$~GPa. Linear fits for $\sigma_{zz} > -2$~GPa yield stress dependencies of $-62$ and $-70$~T/GPa for the $\alpha_{1,2}$-, $124$ and $22$~T/GPa for the $\beta_{1,2}$-, and $-59$, $-76$, and $-74$~T/GPa for the $\gamma_{1,2,3}$-frequencies, respectively. While the absolute slopes differ from those extracted from our measurements, the qualitative evolution is consistent: the $\alpha$ and $\gamma$ frequencies increase, whereas the $\beta$ frequencies decrease with increasing compressive stress. Furthermore, both approaches capture a pronounced enhancement of the $\beta$-sheet warping under $c$-axis stress. In contrast, the $\gamma_3$ frequency has not been experimentally resolved. Since the DFT frequencies are very close to each other, the corresponding experimental peak may be hidden within the $\gamma_{1,2}$ peaks and could be resolved in future high-field measurements.

\begin{table}
    \centering
    \begin{tabular}{c|c|c|c}
    %\multicolumn{4}{c}{Normalised Stress Evolution $\Lambda_n$}\\
        %& \multicolumn{3}{c}{}\\ 
        Peak & dHvA [T/GPa] & dHvA $\Lambda_n$ &  Dilatometer $\Lambda_n$ \\ % & DFT\footnote{We use average values for $\alpha$ and $\gamma$.} \\
        %& 8T-15T& Measurem dHvA [T/GPa] ent \\
        \hline
        $\alpha$ & $-40 \pm 4$ & $1$\footnote{The $\alpha$ sheet was determined in the range $3-15$~T.} & $1$ \\ %& $1$ \\
        $2\alpha$ & $-77\pm 20$\footnote{Fit only for $\sigma_{zz} > -0.4$} & $1.9\pm 0.5$ &  $2.1\pm0.3$ \\% & $-$ \\ 
        
        $3\alpha$ & $-136 \pm 18$ & $3.4\pm 0.5$ & $2.7\pm0.7$ \\% & $-$\\
        $\beta_1$ & $132 \pm 18$ & $3.30\pm 0.5$ & $3.3\pm0.5$ \\ %& $1.88$ \\
        $\beta_2$ & $86 \pm 17$ & $2.2\pm 0.4$ & $2.6\pm0.4$ \\ %& $0.33$ \\
        $\gamma$ & $-84 \pm 19$ & $2.10\pm 0.5$\footnote{The $\gamma$ slope is an average of $\gamma_1$ and $\gamma_2$.} & $3.2\pm 0.7$\footnote{The $\gamma$ sheet was determined in the range $13-15$~T.}\\ % & $1.06$ \\
    \end{tabular}
\caption{Stress evolution of the Fermi sheets with respect to zero stress. The first column lists $dF_n/d\sigma$ from dHvA, while the second and third columns give the stress sensitivities $\Lambda_n$ normalized to the $\alpha$ sheet, as obtained from dHvA and from dilatometer measurements ($8$--$15$~T range).}
    \label{tbl:predictions}
\end{table}

%\footnote{The electron count $N_e = 2\cdot k_\text{F}^2a^2/4\pi$ can be calculate by the Onsager relation $A = 2\pi eF/\hbar$, where $k_\text{F} = \sqrt{A/\pi}$ is an average Fermi wave vector \cite{Mackenzie96_PRL}.}

%%%D
%%% Luttinger volume/charge transfer and rigid band shift
The evolution of the extremal frequencies determined here should preserve the Luttinger count, as required by charge conservation, providing a natural consistency check. Calculating the Luttinger count using the dominant $\alpha_1$, $\beta_2$, and $\gamma_1$ frequencies identified in Fig.~\ref{fig:raw-QO}(c--e) yields approximately 1.783, 0.909, and 1.330 at zero stress, respectively, changing to 1.776, 0.895, and 1.350 at $-1.8$~GPa. The total electron count remains essentially constant, with $n_\mathrm{tot} \approx 4.02$ within experimental uncertainty. We attribute this redistribution of carriers to charge transfer from the Ru $d_{xz}$/$d_{yz}$ orbitals, which mainly form the $\alpha$ and $\beta$ sheets, to the Ru $d_{xy}$ orbital, which mainly forms the $\gamma$ sheet. This interpretation is consistent with ARPES studies of epitaxially strained or doped \SRO{}, which show a redistribution of carriers between these orbitals under tuning of the electronic structure~\cite{Burganov16_PRL,Shen07_PRL}.

In contrast to the charge-transfer mechanism identified here, under $\langle 100 \rangle$ uniaxial stress the Lifshitz transition is reached via a deformation of the $\gamma$ sheet \cite{Sunko19_npj}. 
These distinct mechanisms may account for the large difference in the corresponding critical stresses, namely $\sigma_\text{crit,a}^\text{exp} \approx -0.7$~GPa and $\sigma_\text{crit,c}^\text{DFT} \sim -5.5$~GPa.

%%% Dimensionality (warping) analysis
Further insight into the three-dimensional character of the electronic structure 
is provided by the evolution of the Fermi-surface warping, which is directly 
reflected in the separation of extremal frequencies, in particular 
between the $\beta_1$ and $\beta_2$ orbits (see Fig.~\ref{fig:main_results}(a)). At zero applied stress, the $\beta$ sheet (mainly formed by the Ru $d_{xz}$ and $d_{yz}$ orbitals) is already the most strongly warped Fermi surface and therefore provides the dominant contribution to the out-of-plane conductivity \cite{Bergemann03_AIP}. The increase in the frequency splitting $\beta_1 - \beta_2$ with compressive stress 
indicates an enhancement of the $\beta$-sheet warping, in agreement with measurements 
of the out-of-plane resistivity \cite{Jerzembeck22_NatComm}, and implies that the 
electronic structure of \SRO{} becomes more three-dimensional under $c$-axis stress. In contrast, the warping of the $\alpha$ sheet (also formed by the Ru $d_{xz}$ and $d_{yz}$ orbitals) and the $\gamma$ sheet (mainly formed by the $d_{xy}$ orbital) is only weakly enhanced, in agreement with DFT calculations.
By comparison, under hydrostatic pressure, the $\alpha$ and $\gamma$ sheets were found 
to become more two-dimensional \cite{Forsythe02_PRL}, while changes in the $\beta$-sheet warping could not be resolved.

%%%%%NEW%%%%%%%%%%
In addition to the evolution of the Fermi surface, the effective masses extracted from the dHvA measurements increase under $c$-axis compression. Within a Fermi liquid description, these masses correspond to quasiparticle masses, which are renormalized with respect to the underlying band masses obtained from DFT (Table~\ref{tab:DFT-masses}).

A key question is whether the enhancement of the measured masses reflects an increase in the interaction strength when approaching the Van Hove singularity. To address this, one must consider the ratio between the measured quasiparticle masses and the corresponding DFT band masses. Our results indicate that, although the absolute values of the effective masses increase under stress, the mass enhancement factor remains approximately constant within the experimental uncertainty.

This implies that the increase of the quasiparticle masses is primarily driven by changes in the underlying electronic structure, rather than by a significant enhancement of electron-electron interactions. This behaviour is consistent with previous thermodynamic studies under uniaxial stress, which also found no evidence of increased interaction strength as the Van Hove singularity is approached.
%%%%%%%%%%%%%%%%

%%%E
%Enhancement of the Sommerfeld coefficient and the DOS
Building on this, we estimate the corresponding change in the Sommerfeld coefficient $\gamma_\text{el}$ \cite{Mackenzie96_PRL}. Although the applied stress slightly increases the three-dimensionality of the electronic structure, a two-dimensional approximation for $\gamma_\text{el}$ remains justified.
Based on the dHvA measurements with the out-of-plane magnetic field direction employed here, the Sommerfeld coefficient is given by \cite{Mackenzie96_PRL}:
\begin{equation}
	\gamma_\text{el}(\varepsilon) = \frac{\pi N_\text{A}k_\text{B}^2a^2(\varepsilon)}{3\hbar^2}\sum_{i=\alpha,\beta,\gamma} m^*_i(\varepsilon),
\end{equation}
where $N_\text{A}$ is the Avogadro constant and $a(\varepsilon)$ is the in-plane lattice constant, which changes by only $\sim$0.1\%. 
Using our measured mass data, we obtain $\gamma_\text{el} = 32.4$~mJ/K$^2$mol at 0~GPa and 35.9~mJ/K$^2$mol at $-1$~GPa. These are underestimated by approximately 20\% due to the systematic thermalisation error described above and in the appendix, but as argued above, most of the error should cancel when their ratio is taken.

A simple scaling argument, $H_\text{c2} \propto (T_\text{c}N(E_\text{F}))^2$ \footnote{This assumes that the $k$-dependent gap function changes only by a constant factor.}, suggests an enhancement of $H_\text{c2}$ under compression. At $\sigma_{zz} = -1$~GPa, $T_\text{c}$ is reduced to $\approx 95\%$ of its zero-stress value, while the Sommerfeld coefficient is enhanced by $\sim 11\%$. Using the enhancement of $\gamma_\text{el}$ as a proxy for the increase in $N(E_\text{F})$, this scaling yields an upper critical field enhanced by $\sim 10\%$, in rough agreement with the results of the direct measurement reported in Ref.~\cite{Jerzembeck22_NatComm}.
An open question is why $T_\text{c}$ is enhanced under in-plane uniaxial stress, but not under out-of-plane stress, despite the density of states increasing in both cases. As discussed above, the Lifshitz transition is approached via charge transfer in the latter, rather than purely by a deformation of the Fermi surface. While this charge transfer also enhances the density of states, it may additionally modify the Coulomb screening, which could contribute to the suppression of $T_\text{c}$ under $c$-axis uniaxial stress. There may, however, be other explanations; this remains an open problem that merits further theoretical investigation.

%%%F
%%% Critical stress estimate
Finally, the qualitative agreement between the experimental results and the DFT calculations of the $\gamma$ sheet allows an estimate of the critical stress of the Lifshitz transition. To improve on the rough estimate of $\sigma_\text{crit}^\text{DFT} = 219~\text{GPa} \cdot \varepsilon^\text{DFT}_\text{crit} \sim 5.5$~GPa, we use the relative change of the $\gamma$-sheet peak frequency with stress.

\begin{figure}
    \centering
    \includegraphics[width=\linewidth]{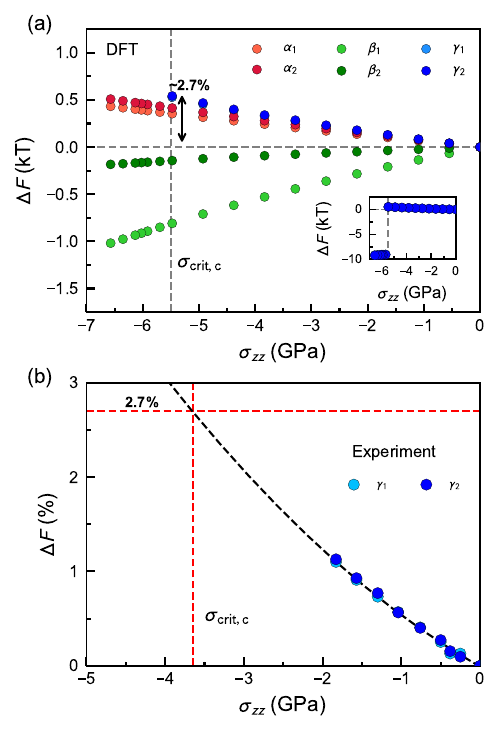}
\caption{(\textbf{a}) Stress dependence of the Fermi-surface cross-sectional 
areas (frequencies) of all sheets calculated by DFT up to $-7$~GPa. 
For the $\gamma$ sheet, a Lifshitz transition occurs at 
$\sigma_\mathrm{crit,c}\!\approx\!-5.5$~GPa, where the Fermi-surface topology 
changes from electron-like to hole-like, as highlighted in the inset. 
The relative change of the $\gamma$ frequency at this point is on average $\sim2.7\%$. 
(\textbf{b}) Relative change of the $\gamma$-sheet frequency from the experimental 
dHvA data. The black dashed line is a second-order polynomial fit; 
by extrapolating this fit to $2.7\%$ we estimate an experimental critical stress 
of $\sigma_\mathrm{crit,c}\!\approx\!-3.7$~GPa.}
    \label{fig:delta_f_sup} 
\end{figure}

Second-order polynomial fits of the DFT-calculated extremal orbit frequencies find that the $\gamma_{1,2,3}$ frequencies change by 2.2, 2.9, and 3.0\% at the critical stress. Taking the average, the DFT critical point corresponds to a relative change of $\approx 2.7\%$.
Since the change of the quantum oscillation frequencies with stress is larger in experiment than in theory, the critical stress at which this relative change is reached is expected to be smaller correspondingly. Using this value as a benchmark, we fit the experimental data for $\sigma_{zz} > -1.85$~GPa with a second-order polynomial and extrapolate the fit to 2.7\%.  This procedure, illustrated in Fig.~\ref{fig:delta_f_sup} (Appendix), yields an estimate of the critical stress of $\sigma_{\mathrm{crit},c} \sim -3.7$~GPa.

A critical stress of $-3.7$~GPa may be experimentally accessible. 
For \SRO{} under in-plane uniaxial stress, DFT was found to overestimate the critical strain by roughly 40\% \cite{Steppke17_Science, Barber19_PRB}, which may also apply to out-of-plane uniaxial stress and thus explain the lower experimental value obtained here. For comparison, ARPES studies under biaxial epitaxial strain in thin films have demonstrated that the Lifshitz transition of the $\gamma$ sheet can be reached by relatively small changes of the in-plane lattice parameter, accompanied by a redistribution of carriers between the Ru $t_{2g}$ orbitals~\cite{Burganov16_PRL}. This highlights that the critical point can be highly sensitive to the specific tuning mechanism, consistent with our observation that $c$-axis stress approaches the transition via charge transfer rather than a simple deformation of the Fermi surface. The maximum applied stress of $-3.2$~GPa achieved in a previous study \cite{Jerzembeck22_NatComm} is already close to this estimate. 
The absence of strong signatures of the Lifshitz transition under $c$-axis stress is not surprising, as both DFT calculations and previous $a$-axis stress experiments show that the Van Hove singularity dominates only in the immediate vicinity of the critical stress \cite{Jerzembeck22_NatComm, Jerzembeck23_PRB}.

\section{Summary}
We report de Haas-van Alphen (dHvA) quantum oscillation measurements under uniaxial stress up to $-1.8$~GPa. 
We find that $c$-axis compression induces a charge transfer from the $\alpha$ and $\beta$ sheets to the $\gamma$ sheet, driving the latter towards a Lifshitz transition and a Van Hove singularity. 
We find good agreement between the stress dependence of the Fermi-surface frequencies determined from dHvA measurements and the magnetostriction oscillations.
Consistent with the approach to a Van Hove singularity, the effective masses of the $\gamma$ Fermi sheets are enhanced, implying an increase of the Sommerfeld coefficient and the density of states. 
These results are in good qualitative agreement with DFT calculations, which capture both the evolution of the Fermi surface frequencies and the enhancement of the effective masses. Comparison of the measured quasiparticle masses with the corresponding DFT band masses shows that the mass enhancement factor remains approximately constant within experimental uncertainty, indicating that the interaction strength does not increase as the Van Hove singularity is approached. 
The agreement between DFT and experiment allows us to estimate the critical stress of the Lifshitz transition as $\sigma_\text{crit,c} \sim -3.7$~GPa, a value that may be experimentally accessible. 
More generally, our results demonstrate that quantum oscillations under uniaxial stress provide a route to study correlated materials at stress values of several GPa. 
It remains an open question how general this approach is and to what extent it can be applied to other correlated materials.

\section{Acknowledgements}  
Research in Dresden benefits from the environment
provided by the DFG Cluster of Excellence ctd.qmat (EXC 2147, Project ID No. 390858490). EH acknowledges funding for the Max Planck Fellowship at the MPI CPfS and for the ERC Consolidator Grant Ixtreme (Grant nr. 101125759). MTP and AWR acknowledge funding from the Engineering and Physical Sciences Research Council [grants EP/P024564/1, and EP/V049410/1] as well as the IMPRS-CPQM.
\section*{Appendix}
\subsection*{Methods}
The dHvA experiments were carried out in Dresden. We studied a single crystal of \SRO{}, grown by a floating-zone method \cite{Bobowski19_CondMat}, with an onset-$T_\text{c}$ of $\approx 1.45$~K.
Figures~\ref{fig:setup}(a-b) show the experimental setup used to measure quantum oscillations under $c$-axis uniaxial stress.
In order to reach large stresses and obtain a high stress-homogeneity, a sample of \SRO{} of 1.85~mm length along the $c$-axis was cut with a Xe focused ion beam into a dumbbell shape (neck diameter 160~$\mu$m x 200~$\mu$m).
This procedure enables higher stresses to be reached \cite{Jerzembeck22_NatComm}.
A pick-up coil of 48 turns (Figure~\ref{fig:setup}(b)) was wound around the neck, whereas the excitation coil consists of 100 turns on a hollow cylinder.
The sample, with both coils around its neck, is glued across a gap of a two-part carrier (Figure~\ref{fig:setup}(a)). 
The two-part carrier prevents the sample from breaking due to differential thermal contraction during cooling and allows easy identification of zero applied stress.
In the last step, the gap carrier is mounted onto a piezo-based uniaxial stress cell \cite{Hicks14_RSI, Barber19_RSI}.
Due to space constraints, we used a uniaxial stress cell with only a displacement (strain) sensor, not a force (stress) sensor.
The uniaxial stress cell is mounted such that the crystallographic $c$-axis is aligned within $3^\circ$ of the cryostat's magnetic field.
Finally, for all quantum oscillation measurements, the ac field was $\sim 4~\mu$T at 71.17~Hz, and we used a low-temperature transformer with an amplification factor of 1000.

\subsection*{Strain-stress conversion}

\begin{figure}[ptb]
\includegraphics{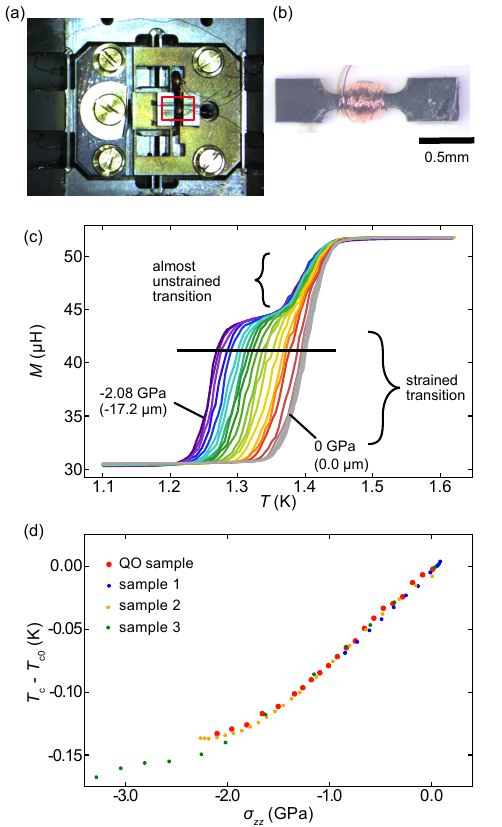}
\caption{(\textbf{a}) Photo of a two-part carrier: the sample is mounted across a gap (red rectangle) with an excitation coil around its neck. (\textbf{b}) $c$-axis sample with a 48 turns pick-up coil. The sample was necked down in the center to reach larger stresses. (\textbf{c}) Mutual inductance against temperature in a series of compressive stresses. We show both the stress and, in brackets, the displacement, applied to the sample. (\textbf{d}) Comparison of $T_\text{c}(\sigma) - T_\text{c}(\sigma = 0)$ against stress. $T_\text{c}$ of the QO sample was determined by a threshold criterion, as indicated in panel (\textbf{c}), and compared to results from Ref.~\cite{Jerzembeck22_NatComm}.}
\label{fig:setup}
\end{figure}

As mentioned above, the uniaxial stress cell used in this experiment was equipped with a displacement (strain) sensor but not with a force (stress) sensor. 
We therefore employed the procedure previously used in Ref.~\cite{Jerzembeck22_NatComm} to convert the measured strain into stress. 
The corresponding stress-strain values used in this work are summarized in Table~\ref{tab:stress_to_strain_exp}. We started by measuring the mutual inductance of the two coils against temperature in a series of compressive strains, which is shown in Figure~\ref{fig:setup}(c).
As in previous experiments, the superconducting transition shifts to lower temperatures with increasing stress.
However, at larger strains, a clear double transition is apparent. 
The double transition occurs because the pick-up coil, which is significantly larger than in previous studies, senses both the strained center of the sample and the almost unstrained ends of the sample.
In a second step, we determine $T_\text{c}$ by using a threshold criterion for the main part of the transition, as indicated by the line in panel \ref{fig:setup}(c).
In a final step, the displacement-to-stress conversion was applied by bringing the rate of change of $T_\text{c}$ over the stress range $-0.92 < \sigma_{zz} < -0.20$~GPa into agreement with the value reported in Ref.~\cite{Jerzembeck22_NatComm}.
Figure~\ref{fig:setup}(d) shows $T_\text{c}$ against stress of the QO sample compared to previous measurements.
Since the samples reported in Ref.~\cite{Jerzembeck22_NatComm} had slightly different zero-pressure critical temperatures, $T_\text{c,0}$, we show the change in $\Delta T_\text{c} = T_\text{c} - T_\text{c,0}$ here.

\begin{table}[h]
\centering
\begin{tabular}{cc}
\hline
$\sigma_{zz}$ (GPa) & $\varepsilon_{zz}$ (\%) \\
\hline
 0.00 &  0.000 \\
-0.25 & -0.114 \\
-0.38 & -0.174 \\
-0.50 & -0.228 \\
-0.76 & -0.347 \\
-1.04 & -0.475 \\
-1.30 & -0.594 \\
-1.57 & -0.717 \\
-1.83 & -0.836 \\
\hline
\end{tabular}
\caption{Experimental $c$-axis stress values converted to strain using $E_{zz}=219$ GPa for Sr$_2$RuO$_4$.}
\label{tab:stress_to_strain_exp}
\end{table}

\subsection*{Mass analysis}

\begin{figure}[ptb]
\includegraphics{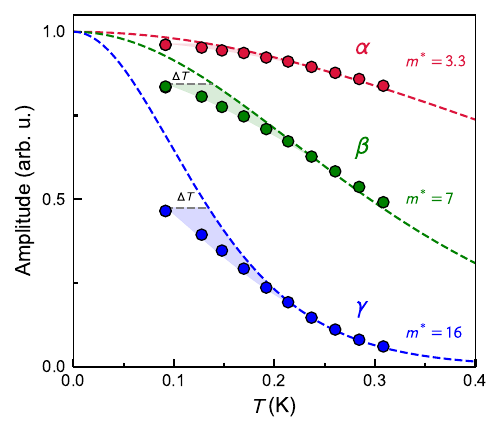}
\caption{Temperature dependence of the FFT amplitudes of the quantum oscillations 
at 0~GPa for the three Fermi sheets ($\alpha$, $\beta$, and $\gamma$). 
Circles represent the experimental data, and dashed lines denote the Lifshitz-Kosevich 
fits using literature-effective masses. The deviation from the expected temperature 
dependence at low $T$ indicates a heating effect. The shaded region illustrates the 
estimated temperature gradient in our setup.}
\label{fig:App_osc}
\end{figure}

The mass analysis for all Fermi sheets was performed in the field range $13$--$15$~T and over the temperature range $90$--$325$~mK. 
We first measured the temperature dependence of the FFT amplitudes with the gap of the two-part carrier open, ensuring zero applied stress. 
Lifshitz-Kosevich (LK) fits yield slightly smaller band masses (by 10--20\%) than previously reported \cite{Mackenzie96_PRL, Bergemann03_AIP}. This indicates a low-temperature heating effect (temperature gradient) between the sample and the thermometer, which becomes apparent below $\sim 200$~mK (see Figure~\ref{fig:App_osc}).  

To avoid introducing model-dependent corrections, we do not apply any temperature correction to the data; all values reported are obtained from direct LK fits to the measured amplitudes (see Figure~\ref{fig:mass_analysis}). As a consistency check, restricting the zero-stress fit to $T > 220$~mK recovers literature masses within uncertainties, as shown in Figure~\ref{fig:App_osc}, where the data follow the LK line.  

For the data at $-1$~GPa, we perform the same fitting procedure (omitting the noisiest points above $250$~mK for the $\gamma$ sheet).  
Even without applying a temperature correction, we find that the band masses increase from $3.0$, $6.0$, and $13.0$ to $3.2$, $6.2$, and $15.0$ for the $\alpha$, $\beta$, and $\gamma$ Fermi sheets, respectively.  
This corresponds to an average enhancement of $\sim$ $10$\% (within the error bars) under $c$-axis compression, directly visible in the raw data.

\subsection*{Dilatometer\label{apn:Dil}}

\begin{figure}
    \centering
    \includegraphics[width=0.75\linewidth]{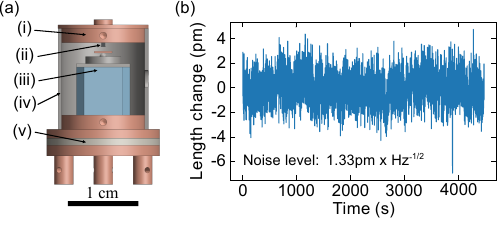}
    \caption{\textbf{(a)} Magnetostriction setup with (i) Lid with passthrough electrode, (ii) Stress-free sample mounting, (iii) Attocube for cryogenic approach and calibration, (iv) Titanium construction matches Attocube and minimises background, (v) PPMS compatible adapter. \textbf{(b)} The noise floor of the dilatometer at $15$~T and base temperature after subtraction of a slowly varying background due to a slight temperature instability.}
    \label{fig:dilatometer}
\end{figure}

Our dilatometry measurements were carried out in St Andrews. In order to measure the strain quantum oscillations we have developed a new capacitive dilatometer based around an Attocube\texttrademark{} linear actuator.
This dilatometor, shown schematically in Figure~\ref{fig:dilatometer}(a), uses the top surface of the sample itself (or a small polished metal film glued on to the face in the case of a rough surface) as a capacitive plate, with the other plate mounted on a ball joint on top of the Attocube\texttrademark{} which is free to move relative to the sample as the Attocube\texttrademark{} expands and contracts.
The Attocube\texttrademark{} can be withdrawn for cooling, ensuring no damage due to thermal contractions.
For measurement, the Attocube\texttrademark{} is expanded as far as possible without shorting towards the sample face and is grounded to achieve the highest possible baseline capacitance (and thus sensitivity).
The sample is then free to expand and contract without resistance in response to changing parameters (for example, field or temperature), with its changing length directly changing the plate separation and thus capacitance of the system.
The relationship between capacitance and length change can be determined from the known extension of the piezoelectric stacks in the Attocube\texttrademark{} in response to small biases and has been found to be linear for the range of sample extensions observed in response to field or temperature changes below 1~K.
The advantage of the method is that it allows working with extremely small capacitor plate separation, and correspondingly high sensitivity. Using our setup, we have achieved a plate separation of approximately $1~\mu$m with a noise floor typically less than 1.5~pm/$\sqrt{\text{Hz}}$ and below 1.0~pm/$\sqrt{\text{Hz}}$ under ideal conditions.
An example of the noise floor measured at base temperature and 15~T is shown in Figure~\ref{fig:dilatometer}(b).

\subsection*{Density functional theory calculations}

The calculations of the Fermi surfaces of \SRO{} were carried out by density functional theory (DFT) calculation in the local density approximation (LDA) with the Perdew-Wang parametrization \cite{Perdew96_PRL} for the exchange-correlation potential.
The calculations were performed by using the full-potential local orbit FPLO version fplo 22.00-52 (http://www.fplo.de) \cite{Koepernik99_PRB, Opahle99_PRB}.
Spin-Orbit coupling was included by using the four-component Kohn-Sham-Dirac equation \cite{Eschrig04}.
In order to obtain precise band structure and Fermi surface information in the vicinity of a Van Hove singularity, the calculations were carried out on a well-converged mesh with 343000 $k$-points ($70 \times 70 \times 70$; 23022 points in the irreducible wedge).
Ambient pressure lattice parameters at 15~K from Ref.~\cite{Chmaissem98_PRB} were used.
Strain was applied by using the out-of-plane strain $\varepsilon_{zz}$ as the independent variable and the in-plane strains ($\varepsilon_{xx}$ and $\varepsilon_{yy}$) were set by the low-temperature Poisson's ratio \cite{Ghosh21_NatPhys}.
The apical oxygen position was relaxed for each strain independently by minimizing the force below 1~meV/$\textup{~\AA}$.
The Fermi surface frequencies were obtained by using the Onsager relation ($A = 2\pi eF/\hbar$) and the extremal cross-sectional areas of the individual Fermi surfaces for each strain. 
The band masses for $\varepsilon_{zz} = 0$ and $-0.01$ were calculated from a Fermi surface average of the $k$-dependent Fermi velocity $v(k)$ of the extremal cross sections and are shown in Table~\ref{tab:DFT-masses}.

\begin{table*}
\centering
%\footnotesize
\begin{tabular}{p{1cm}|p{1cm}|p{1cm}|p{1cm}|p{1cm}|p{1cm}|p{1cm}|p{1cm}|p{1cm}|p{1cm}}
$\varepsilon_{zz}$  & $\alpha_1$ & $\alpha_2$ & $\alpha_3$ & $\beta_1$ & $\beta_2$ & $\gamma_1$ & $\gamma_2$ & $\gamma_3$ & $\gamma_4$ \\
\hline
 $0.0$ & $1.06$ & $1.06$ & $1.05$ & $1.88$ & $2.07$ & $3.25$	 & $3.25$ & $3.33$ & $3.34$ \\ 
 $-0.01$ & $1.11$ & $1.11$ & $1.10$ & $1.93$ & $2.17$ & $3.57$ & $3.57$ & $3.70$ & $3.72$  
\end{tabular}
\caption{DFT band masses magnitude for the nine extremal orbits at $\varepsilon_{zz} = 0$ and $-0.01$.}
\label{tab:DFT-masses}
\end{table*}

%%%%%%%%%%%%%%%%%%%%%%%%%%

%%%%%%%%%%%%%%%%%%%%%%%%%%%%%%%%%%%%
%\section{Data Availability}

%Raw data will be made available in a public repository upon publication.

\bibliography{bibliography.bib}
\bibliographystyle{apsrev4-2}

\end{document}